# Lévy Flight of Holes in InP Semiconductor Scintillator

**Serge Luryi and Arsen Subashiev**
*Dept. of Electrical and Computer Engineering, State University of New York
Stony Brook, NY 11794-2350, U.S.A.*

## 1. Introduction

The term "Lévy flight" (LF) was coined by Benoît Mandelbrot to describe a random walk, in which step lengths $\ell$ have a probability distribution that is heavy-tailed. Although the exact definitions of "heavy tailing" vary in the literature, we shall reserve the term to distributions $\mathcal{P}(\ell)$ that do not possess a variance as they decrease too slowly at large steps, $\mathcal{P}(\ell) \propto \ell^{-(1+\gamma)}$. For the index $\gamma$ in the range $0 < \gamma < 2$, the distribution itself can be normalized, $\int \mathcal{P}(\ell)\,d\ell = 1$, but its second moment, $\langle \ell^2 \rangle = \int \ell^2 \mathcal{P}(\ell)\,d\ell$, diverges. Although one often speaks of "anomalous diffusion", the LF random walk cannot be described by an ordinary diffusion equation. The conventional diffusivity is not even defined for such a random walk.

Possibility of statistical description of the random walk (e.g., through evaluation of a particle distribution that emerges from a point-type source after a given number of steps) relies on a statistical theorem that defines the limit of a sum of randomly distributed numbers (in our case these are the lengths of individual steps). If the step length distribution $\mathcal{P}(\ell)$ decreases rapidly enough for large steps (namely, when $\gamma > 2$) the result is given by the Central Limit Theorem (CLT) and the sum has a normal (Gaussian) distribution. When the steps are distributed with heavy tails, their sum does not follow the CLT and is not Gaussian. It may be still described by a universal (though $\gamma$-dependent) distribution, called the stable distribution. The first systematic studies of the stable distributions originate from Paul Lévy and Aleksandr Khinchin [1].

The LF transport problem has been extensively studied mathematically. Description of the anomalous transport in terms of fractional dynamic equations or, for random walks in external field, fractional Fokker-Planck equations, is amply discussed in the reviews [2-4]. These phenomena are well-known to astrophysicists, as they occur in the problem of transport of resonance radiation in celestial bodies [5, 6]. They are also known in plasma physics as the imprisonment of resonance radiation in gaseous discharge [7, 8]. Interestingly, LF transport is more common in nature than one might think: thus, Lévy flights were recently invoked to explain movement strategies in mussels as revealed in the patterning of mussel beds [9], as well as ocean predators search strategies in regions where prey is sparse [10]. Birds and other animals also seem to follow Lévy flights when foraging [11]. Finally, a vast literature is devoted to Lévy flights in finance, "random walk down the Street" [12].

Nevertheless, there have been preciously few experimentally available laboratory systems for studying LF transport, ideally with variable parameters. A rather ingenious such system was recently demonstrated by Barthelemy *et al.* [13], who embedded scattering particles in a glass matrix – together with non-scattering glass microspheres of same refractive index as the matrix. The sole purpose of these spacer spheres was to modify locally the average separation between the scattering particles and thus control the step-length distribution for photon transport. With specially designed, highly non-trivial, distributions of microspheres diameter, the authors were able to observe a Lévy flight of light.

Recently, we described [14] a more "natural" lab system exhibiting Lévy flight, namely the direct-gap semiconductor of high radiative efficiency, specifically *n*-doped InP. The randomly walking particles in this case are minority carriers (holes) and their dominant transport process is photon-assisted hopping. This process, also known as the photon recycling, consists of radiative recombination of a hole at one spot producing a photon, whose subsequent interband absorption leads to the re-emergence of a hole at another spot, possibly far away. The high radiative efficiency and low free-carrier absorption of light in lightly doped InP ensure that photon recycling continues for about 100 times before a hole recombines non-radiatively or a photon is absorbed without leaving a hole behind. The randomness of free flight is set by the emission spectrum in radiative recombination. This spectrum, combined with the interband absorption probability and the probability of photon propagation to a given distance, defines the probability distribution for free flights of photons. Photons generated in the long-wavelength wing of the emission spectrum travel long distances before they get re-absorbed and are responsible for the divergent variance of the distribution and the Lévy-flight nature of the resulting random walk. This process is reviewed in Sect. 2.

Manifestations of anomalous transport were found [14] by studying photoluminescence in *n*-doped InP. The key evidence was derived from the ratio of transmitted and reflected luminescence spectra, measured in samples of the same doping level but very different thicknesses (350 μm vs. 50 μm). The results give a direct experimental proof of the non-exponential decay of the minority-carrier concentration from the surface where the holes were photo-excited initially. The power-law decay of the hole concentration, characteristic of the LF transport, is steep enough at short distances (steeper than an exponent) to fit the data for the thin sample, and at the same time slow enough at large distances (again, compared to an exponent) to account for the data for thick samples. This work is reviewed in Sect. 3.

Transport at much larger distances (up to centimeters) was studied in experiments [15], where photoluminescence was registered from the edge of an InP wafer as a function of the distance from the excitation spot on the broadside surface. Since the extremely long photon propagation is owing to the transparency region at the red wing of the emission spectrum, one observes a red shift in the luminescence spectrum, with larger shift corresponding to longer distances. Analysis of this shift provides an independent and accurate determination of the Urbach tails in moderately doped semiconductors. This work is reviewed in Sect. 4.

Sections 5 and 6 deal with practical applications of the anomalous transport of minority carriers in semiconductors of high radiative efficiency, specifically to the so-called semiconductor scintillator [16-18]. Normally, scintillators are not made of semiconductor material. The key issue in implementing a semiconductor scintillator is how to make the material transmit its own infrared luminescence, so that the response signal generated deep inside the semiconductor slab could reach its surface without tangible attenuation. In high-efficiency semiconductors, the long tails of Lévy-flight transport come to the rescue, providing near-ideal photon collection. Luminescence experiments [14, 15] support a simple model of photon collection, which we shall refer to as the "on the spot approximation" (OTSA). In this model, the signal received by a photodetector at the surface arises from repeated emission at the same spot where the initial minority carrier was generated. Each attempt has a small probability of "unhappy" termination, due to the nonradiative channel of recombination or free-carrier absorption. The happy end corresponds to the photon reaching the surface and being collected at the photodetector. The advantage of the OTSA is that it leads to a close-form expression for the collected signal, by summing a geometric series. As discussed in Sect. 5, the OTSA is very close to reality for the typical minority-carrier distributions generated by Lévy-flight transport.

Our understanding of the anomalous transport of minority carriers in direct-gap semiconductor of high radiative efficiency has led to the invention [18] of a layered scintillator, described in Sect. 6. The idea of embedding radiation sites in a semiconductor (or insulator) material is nearly as old as the scintillator concept itself [19]. In all such devices [20], the photo-generated carriers migrate to the radiation sites and recombine there emitting deep subband light, for which the material is transparent. The density of the radiation sites must be very high, so that the typical travel distance for carriers is much smaller than their diffusion length. The novelty of our idea [18] is to employ the photon-assisted transport of minority carriers rather than their ordinary diffusion. This allows one to space out the radiation sites (in our case, narrow low-bandgap wells embedded in a wide-gap semiconductor matrix) by a large distance. Ultimately, this may lead to the implementation of centimeter-thick semiconductor scintillators.

Our conclusions will be summarized in Sect. 7.

## 2. Photon assisted random walk of minority carriers in InP

Suppose that an electron-hole pair is created by optical excitation in an *n*-doped infinite crystal. There is no interest in tracing the additional single electron as it produces little change in the majority-carrier system. All the action is due to the additional hole. Firstly (on a sub-picosecond time scale), it will become "thermalized", i.e. lose the excess energy it obtained from the light quantum. On a longer (nanosecond) time scale, the hole will move randomly with the thermal velocity until recombining with an electron. This type of random walk corresponds to the ordinary diffusion. The average hole lifetime $\tau$ depends on the electron concentration and is in the nanosecond range. The recombination process can be either radiative or non-radiative, and the rates of these processes are additive, $\tau^{-1} = \tau_{rad}^{-1} + \tau_{nr}^{-1}$. The probability of radiative recombination is described by the emission quantum efficiency $\eta$, viz.

$$\eta = \frac{\tau_{nr}}{\tau_{nr} + \tau_{rad}}. \tag{2.1}$$

The non-radiative lifetime in high-quality crystals reaches several microseconds, resulting in $\eta > 90\%$. The emitted photons disappear mainly via interband absorption process, resulting in the generation of a new hole and then a new photon emitted via radiative recombination. The absorption-reemission sequence will be repeated many times until the recycled hole recombines non-radiatively or the intermediate photon is destroyed by a residual non-interband absorption process. This sequential process is called the photon recycling. Due to the short thermalization time of holes, the emission spectrum remains the same at all stages of the recycling and is well described by the equilibrium electron-hole recombination spectrum.

## *2a. Diffusion equation with a recycling term.*

The spatial distribution of holes is formed by two additive transport processes: (*i*) the random flights of holes (at sub-micron distances) interrupted by scattering, as in the ordinary diffusion, and (*ii*) the photon-assisted transfer of holes over much larger distances. To quantify these processes we write down a modified diffusion equation for the concentration of holes $p(\mathbf{r},t)$:

$$\frac{\partial p}{\partial t} + D\Delta p = -\frac{p}{\tau} + G(\mathbf{r},t) + R(\mathbf{r},t), \tag{2.2}$$

where $D$ is the diffusivity of holes, $\tau$ is the hole lifetime against all recombination processes, and $G(\mathbf{r},t)$ is the generation function defined as the concentration of holes generated per unit time. For a single hole generated at $\mathbf{r} = 0$ and $t = 0$, this function is $G(\mathbf{r},t) = \delta(\mathbf{r})\delta(t)$. The last term $R(\mathbf{r},t)$ is the recycling function,

$$R(\mathbf{r},t) = \frac{\eta}{\tau} \int P(|\mathbf{r}-\mathbf{r}'|) p(\mathbf{r}',t) \, d\mathbf{r}', \tag{2.3}$$

which describes the concentration of holes generated per unit time at point $\mathbf{r}$ due to the radiative recombination of holes present in the crystal at the time *t*.

The factor $P(|\mathbf{r}-\mathbf{r}'|)$ in the integrand of Eq. (2.3) describes the probability that a hole at $\mathbf{r}'$ generates another hole at $\mathbf{r}$ by the described emission-reabsorption process. For the two points separated by the distance $r=|\mathbf{r}-\mathbf{r}'|$, this probability is given by

$$P(r) = \int \mathcal{N}(E) \frac{\exp[-\alpha_i(E)r]}{4\pi r^2} \alpha_i(E)\, dE, \qquad (2.4)$$

where $\alpha_i(E)$ is the light absorption coefficient due to interband processes only. The integrand in Eq. (2.4) is the product of three probabilities: (*i*) the probability of emission of a photon of energy *E*, described by the normalized emission spectral function $\mathcal{N}(E)$; (*ii*) the propagation probability of this photon over the distance $r=|\mathbf{r}-\mathbf{r}'|$ (this probability is described by the intensity distribution produced by a unit point source) ; the absorption probability of this photon, described by the factor $\alpha_i(E)$. One can easily check that the probability (2.4) is properly normalized, $\int P(r)\,dV = 1$.

Equation (2.2) with the recycling term (2.3) was first obtained in the papers by Holstein [7] and Biberman [21] for the radiative spread of the excited-atom concentration in gases and is known as the Biberman-Holstein equation. Solution of this equation in 3D case is complicated by the fact that the resultant distribution cannot be factorized as a product of distributions along perpendicular axes. In other words, in contrast to the familiar Gaussian distribution, projections of the displacements on coordinate axes are correlated.[1] The source of these correlations and the entire difficulty reside in the recycling term. Nevertheless, one can find the solution of Eq. (2.2) quite generally by a trick due to V. A. Ambartsumyan [22], which transforms Eq. (2.2) into an equation for $\tilde{p}(z,t)$, which is the concentration $p(\mathbf{r},t)$ integrated over the $(x,y)$ plane:

$$\tilde{p}(z,t) \equiv \int p(\mathbf{r},t)\,dxdy = 2\pi \int_z^\infty p(z,\rho,t)\rho\, d\rho, \qquad (2.5)$$

---

[1] This can be illustrated in the instance when the characteristic function $F(\mathbf{k})$ of the distribution is of the form

$$F(\mathbf{k}) \equiv \int P(\mathbf{r})\exp[-i\mathbf{k}\cdot\mathbf{r}]\,dV = r_0|\mathbf{k}|$$

In the 1-dimensional case, the inverse Fourier transformation of $F(k_i)$ generates a Cauchy distribution,

$$P_1(x_i) = \frac{1}{\pi}\frac{r_0}{x_i^2 + r_0^2}$$

In the *d*-dimensional case, the transform of $F(\mathbf{k})$ yields

$$P_d(r) = \frac{\Gamma[(1+d)/2]}{\pi}\frac{r_0}{\left(r^2+r_0^2\right)^{(1+d)/2}}$$

where $r^2 = \sum x_i^2$. It is evident that the above expression for $P_d(r)$ cannot be factorized, and hence different $x_i$ components are manifestly correlated.

With the known $\tilde{p}(z,t)$, one can find $p(r,t)$ by differentiating Eq. (2.5) with respect to $z$,

$$p(r,t) = -\frac{1}{2\pi r}\frac{\partial \tilde{p}(z,t)}{\partial z}\bigg|_{z=r}, \tag{2.6}$$

The 1D concentration $\tilde{p}(z,t)$ obeys a much simpler equation that is obtained by integrating Eq. (2.2) over the $(x,y)$ plane. The resulting 1D equation is of the form [23]

$$\frac{\partial \tilde{p}}{\partial t} - D\frac{\partial^2 \tilde{p}}{\partial z^2} + \frac{\tilde{p}(z,t)}{\tau} = \frac{\eta}{2\tau}\int_{-\infty}^{\infty}\tilde{p}(z',t)\,\mathcal{P}(|z-z'|)\,dz' + \tilde{G}(z,t), \tag{2.7}$$

where $\tilde{G}(z,t)$ is the generation term $G(\mathbf{r},t)$ integrated over the $(x,y)$ plane. The probability $\mathcal{P}(|z-z'|)$ is given by

$$\mathcal{P}(z) = \int \mathcal{N}(E)\,\alpha_i(E)\,\mathrm{Ei}[1,\alpha(E)z]\,dE, \tag{2.8}$$

where $\mathrm{Ei}(1,z)$ is the exponential integral function.[2] The probability $\mathcal{P}(|z-z'|)$ satisfies the normalization condition

$$\mathcal{P}_{\mathrm{tot}} \equiv \int_0^{\infty}\mathcal{P}(z)\,dz = 1. \tag{2.9}$$

If one knows $\mathcal{P}(|z-z'|)$, then, for an infinite medium, Eq. (2.7) can be solved [24] by a Fourier transformation. Using this equation, one can study the temporary evolution of the total number of holes per unit length along the $z$ axis. In view of its linearity, Eq. (2.7) can be equally well applied to the case of a planar excitation uniform in the $(x,y)$ plane – with $\tilde{G}(z,t)$ being the hole concentration generated per unit time. In this case, $\tilde{p}(z,t)$ is just the $z$-dependent concentration of holes.

In the problem of interest to us, the ordinary diffusion term gives negligible contribution. With this term dropped (by setting $D=0$), Eq. (2.7) retains a simple probabilistic interpretation: it describes a 1-dimensional random walk of a particle created at $z=0$. The distribution of jump lengths is given by $\mathcal{P}(|z|)$, the average time between jumps is $\tau$, and $1-\eta$ is the probability of particle loss at any step. This interpretation suggests that Monte Carlo modeling should be a useful approach to studying $\tilde{p}(z,t)$. It has the advantage of being able to include various factors "difficult" in any analytic approach, such as effects of the boundaries, realistic shape of the generation pulse, etc. Our calculation begins with ascertaining the distribution $\mathcal{P}(|z|)$ of single jumps in the photon-assisted random walk of holes in $n$-InP.

---

[2] This function is defined by $\mathrm{Ei}(1,z) = \int_1^{\infty} t^{-1}\exp(-zt)\,dt$

## *2b. Jump distribution.*

We shall use Eq. (2.8) to evaluate $\mathcal{P}(|z|)$ from the experimentally measured [15, 25] interband absorption coefficient $\alpha_i(E)$ for moderately doped *n*-type InP. With the known $\alpha_i(E)$, the spectral density $\mathcal{N}(E)$ of photon emission in the quasi-equilibrium hole recombination process (minority holes recombining with majority electrons) can be obtained by the thermodynamic relation due to van Roosbroek and Shockley [26],

$$\mathcal{N}(E) = A\alpha_i(E) E^2 e^{-E/kT}, \tag{2.10}$$

which we shall refer to as the VRS relation. Expression (2.10) represents the "intrinsic" emission spectrum and it agrees very well with the spectra of luminescence measured [27] from thin epitaxial layers (especially when those are clad by wider-gap layers to prevent surface recombination). Experimentally observed bulk luminescence spectra differ from VRS and the distortion depends on the geometry of the experiment. As discussed in Sects. 3 and 4, the main spectral distortion results from energy-dependent filtering due to the re-absorption of outgoing photons. These experimentally accessible filtering functions contain a wealth of information about the steady-state minority-carrier distribution $\widetilde{p}(z)$. In this section, concerned with evaluation of $\mathcal{P}(|z|)$, we are not interested in filtering and assume that the intrinsic (unfiltered) emission lineshape is faithfully given by the VRS relation.

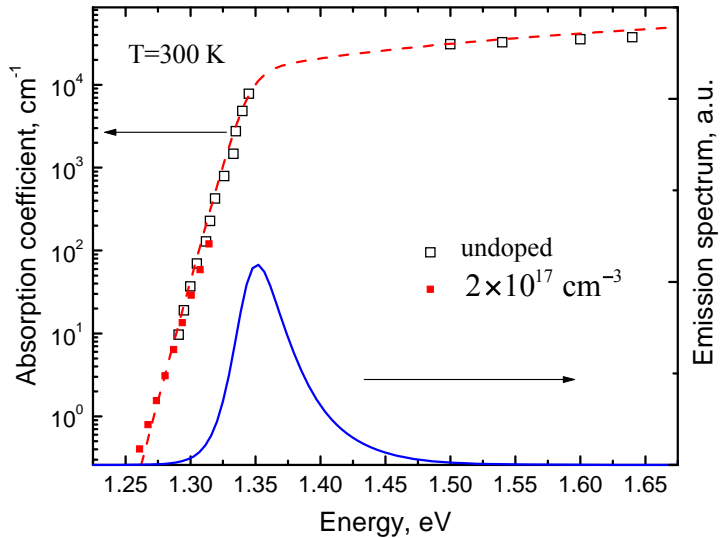

**Fig. 2.1.** Experimentally observed absorption spectrum (log scale) for moderately doped *n*-InP sample ($N_\mathrm{D} = 2 \times 10^{17}$ cm$^{-3}$). Dashed lines show the fitting by Eq. (2.12). The intrinsic emission spectrum, derived from the VRS relation (2.10), is plotted on the linear scale.

Both spectra, $\alpha_i(E)$ and $\mathcal{N}(E)$, are displayed in Fig. 2.1. Below the absorption edge, $\alpha_i(E)$ decreases exponentially

$$\ln \frac{\alpha_i(E)}{\alpha_i(E_G)} = \frac{E - E_G}{\Delta}, \qquad (2.11)$$

If the bandgap $E_G$ is independently known, the $\alpha_i(E)$ dependence is characterized by two parameters, absorption at the bandgap, $\alpha_i(E_G)$, and the tailing energy $\Delta$ (Urbach tail). In a broader range that includes the absorption edge and the region $E > E_G$ the dependence $\alpha_i(E)$ for moderately doped samples ($N_D < 10^{18}$ cm$^{-3}$) is well approximated by

$$\alpha_i(E) = \frac{E - E_0}{E_G} \frac{\alpha_0}{1 + \exp[(E_G - E)/\Delta]}, \qquad (2.12)$$

see the dashed line in Fig. 2.1. Here the first factor reflects an almost linear growth of $\alpha_i(E)$ above the absorption edge and the second factor reproduces Urbach tailing (2.11). For the electron concentration $N_D = 2 \times 10^{17}$ cm$^{-3}$, parameters in (2.12) are $E_G = 1.35$ eV, $E_0 = 0.9$ eV, $\Delta = 8.2$ meV, and $\alpha_0 = 6.6 \times 10^5$ cm$^{-1}$. For higher concentrations, an increase is observed [15] in both $E_G$ and $\Delta$ (for $N_D = 2 \times 10^{18}$ cm$^{-3}$ we have $E_G = 1.36$ eV, $E_0 = 1.2$ eV, $\Delta = 10.6$ meV, and $\alpha_0 = 1.2 \times 10^5$ cm$^{-1}$). Note that the Fermi level crosses $E_G$ at $N_D = 4.3 \times 10^{17}$ cm$^{-3}$ (at room temperature). At higher concentrations, the absorption spectra are influenced by the conduction band filling (the Moss-Burstein shift).

Results of numerical evaluations of $\mathcal{P}(|z|)$ with Eq. (2.8) are shown in Figs. 2.2 (a, b). In the entire range of $z$, the dependence is very close to

$$\mathcal{P}(z) = \frac{\gamma \, z_{\min}^{\gamma}}{2 (z_{\min} + z)^{1+\gamma}}, \qquad (2.13)$$

where $z_{\min} \approx 0.1 \, \mu$m is a non-essential parameter describing the short-distance behavior. The essential parameter is the exponent $\gamma$, called the index of the distribution. For moderately doped samples, in the range $N_D = (2 \text{ to } 6) \times 10^{17}$ cm$^{-3}$ illustrated in Fig. 2.2(a), the index varies from $\gamma \approx 0.69$ to 0.64, slightly decreasing with the doping level.

Theoretical calculation of the index $\gamma$ in an analytical form is possible in a model, where $\alpha_i(E)$ is approximated by a function simpler than (2.12):

$$\alpha_i(E) = \frac{\alpha_0}{1 + \exp[(E_G - E)/\Delta]}. \qquad (2.14)$$

The function (2.14) decays exponentially below the absorption edge and saturates above it. This model correctly accounts for the Urbach tailing but it does not describe the approximately linear growth of $\alpha_i(E)$ at $E > E_G$. Furthermore, the emission spectrum in this model can be described by Eq. (2.10) with $\alpha_i(E)$ given by (2.14) and the pre-exponential factor replaced by its value at $E = E_G$ (reasonable for $\Delta \leq kT \ll E_G$). The model yields a simple expression for the index,

$$\gamma = 1 - \frac{\Delta}{kT}. \tag{2.15}$$

Equation (2.15) predicts lower values of $\gamma$ at lower temperatures. It also explains the decrease of $\gamma$ with increasing $N_D$. The latter effect is due to the smearing of the absorption edge at higher doping, described by increasing tailing energy $\Delta$. Estimation of the index with Eq. (2.15) for moderately doped samples ($\Delta = 9.4$ meV) gives $\gamma \approx 0.64$ in close agreement with the results obtained by more accurate numerical calculations.

For heavier doping, the accurate numerical calculations yield a more complicated concentration dependence of $\gamma$ (see Table I) compared to that predicted by (2.15). The discrepancy is due to the fact that Eq. (2.14) is no longer a good approximation when the Moss-Burstein shift is large.

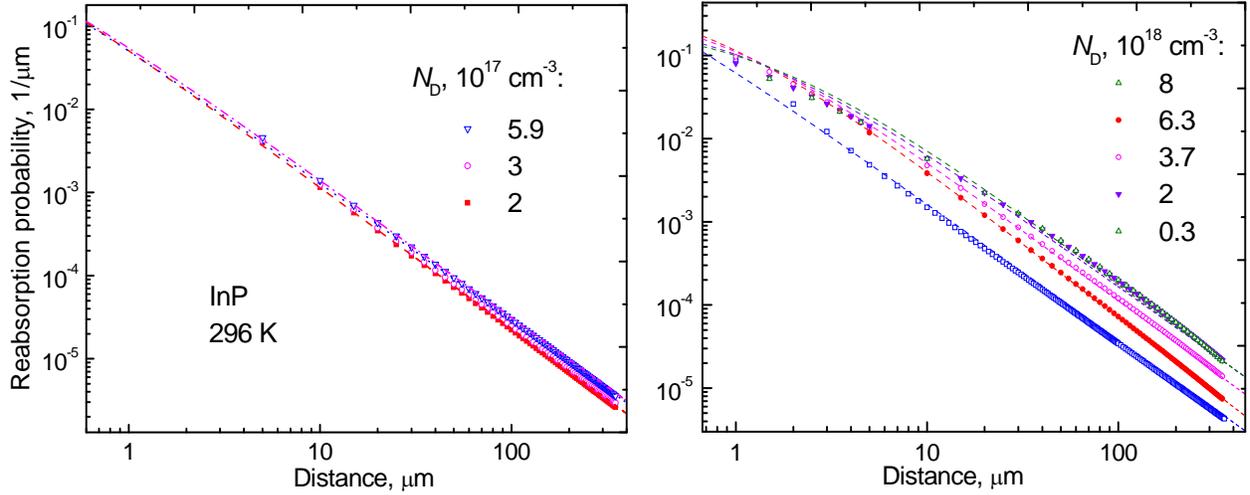

**Fig. 2.2.** Reabsorption probability $\mathcal{P}(z)$ calculated with Eq. (2.8) for moderately doped (a) and heavily doped (b) samples at room temperature. Relevant sample parameters are listed in Table I. Dashed lines correspond to the power-law approximation, Eq. (2.13). Note that the $3 \times 10^{17}$ cm$^{-3}$ sample is represented in both graphs for comparison.

Finally, we note that in addition to interband absorption, *n*-doped InP has a residual "free-carrier" absorption, $\alpha_{fc}(E)$, which linearly grows with the doping and weakly depends on the energy in the vicinity of the interband absorption edge. It can be easily taken into account by replacing $\alpha_i(E) \to \alpha(E) = \alpha_i(E) + \alpha_{fc}(E)$ in the exponential propagation probability factor in Eq. (2.4). In Eq. (2.8), similar replacement has been actually done tacitly in the argument of the Ei function. As a result, the full probability (2.9) of interband reabsorption becomes less than unity, $\mathcal{P}_{tot} < 1$. However, this effect is small. Experimentally, $\alpha_{fc} = 0.13 \times 10^{-17} N_D$ cm$^{-1}$ at room temperature, and in the range of distances of our interest the effect of free-carrier absorption is negligible for all experiments discussed below.

## *2c. Stable distribution of minority carriers.*

The emission-reabsorption probability (2.8) describes a one-dimensional photon-assisted motion of holes that we call "jumps" in distinction from the actual hole movement interrupted by scattering and described as conventional diffusion. The stochastic nature of this 1D random walk is associated with the emission spectrum $\mathcal{N}(E)$. The hole jump probability, accurately described by Eq. (2.13), is typical for the "anomalous diffusion" of the Lévy-flight type [3]. Its hallmark is the asymptotic spatial decay with a "heavy tail," $\mathcal{P}(z) \propto 1/z^{1+\gamma}$, which has the power-law asymptotic index $\gamma < 2$. Since the second moment of this distribution diverges, one cannot describe the photon-assisted random walk of holes with a conventionally defined "enhanced" diffusion coefficient $D_{enh} \propto \langle z^2 \rangle / \tau$. However, for any finite number *N* of jumps $z_i$, one can find the statistical (averaged over many histories) hole distribution $p(z, N)$ for $z = \sum_{i=1}^{N} z_i$.

Monte Carlo simulation is well-suited for this purpose. The normalized hole distributions simulated with $\gamma = 0.76$ are presented in Fig. 2.3(a) for several values of *N*. The initial excitation is localized at the origin. The heavy tails are very prominent on the logarithmic scale, especially in comparison with similar distributions shown in Fig. 2.3(b) for the assumed index $\gamma > 2$, when $\langle z^2 \rangle$ is finite and the random walk is Brownian.

The statistical model of a random walk on an infinite line has been thoroughly studied [2, 3, 28]. After $N \gg 1$ jumps originating from $z = 0$, the distribution approaches the so-called *stable distribution* [28] with a given index $\gamma < 2$, viz.

$$p(z, N) = \int_0^\infty \cos(kz) \exp\left[-N(z_C k)^\gamma\right] dk, \qquad (2.16)$$

where $z_C$ is a depth scaling factor, which depends on the short distance behavior of the jump distribution $\mathcal{P}(z)$. In our case, $z_C \approx z_{min}$ by the order of magnitude. Comparing the distribution (2.16) with our Monte Carlo results, we find $z_C = 0.23\,\mu\text{m}$. Figure 2.3(a) displays both the results of Monte Carlo simulations and calculation using Eq. (2.16). Excellent agreement demonstrates high accuracy of the Monte Carlo approach that will be extended below to include "difficult" factors, such as various boundary conditions in the random walk over a finite slab.

According to Eq. (2.16), the stable distribution is of the form

$$p(z, N) = \frac{1}{N^{1/\gamma}} g_\gamma \left( \frac{1}{N^{1/\gamma}} \frac{z}{z_C} \right), \qquad (2.17)$$

with the universal function $g_\gamma(z)$ characterized by only one parameter $\gamma$. For this reason, the distribution is called "strictly stable", its universality being similar to the universally Gaussian shape of the normal distribution that emerges in the case of $\gamma > 2$, when the random walk is Brownian. Figure 2.3 (b) illustrates the Monte Carlo simulated distributions assuming $\mathcal{P}(z)$ of the form (2.13) with the index $\gamma = 3.5$. In this case, the 2$^{nd}$ moment is finite, $\langle z^2 \rangle = (8/15) z_{min}^2$, and for large $N$, the distributions $p(z, N)$ are normal, according to the central limit theorem.

An important conclusion can be drawn by examining the asymptotic behavior of $g_\gamma(z)$. It follows from (2.16) that at large $z$

$$g_\gamma(z) \big|_{z \gg z_C} \to \left( \frac{z_C}{z} \right)^{1+\gamma}, \qquad (2.18)$$

which implies that

$$p(z, N) \big|_{z \gg z_C} \to N \left( \frac{z_C}{z} \right)^{1+\gamma}. \qquad (2.19)$$

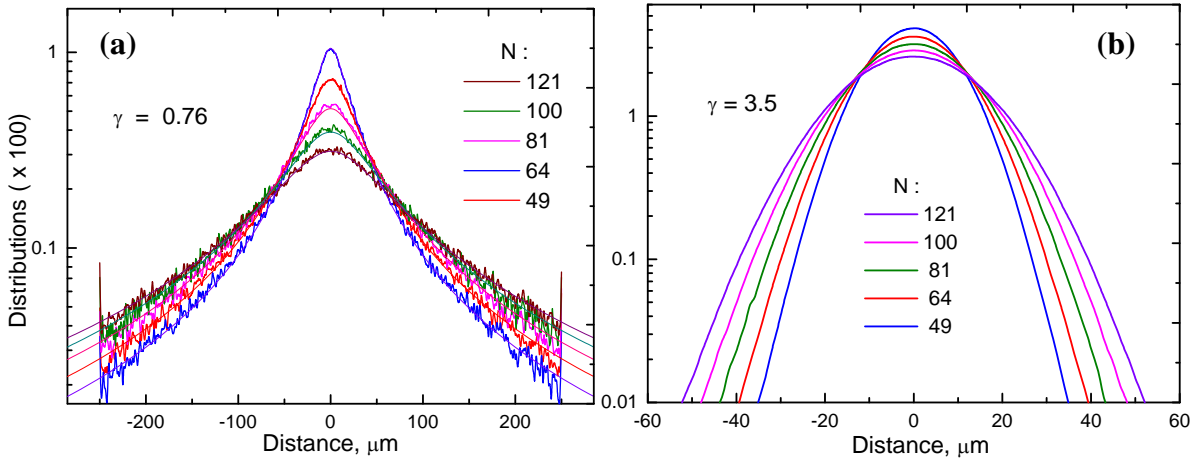

**Fig. 2.3.** (a) Hole distribution $p(z, N)$ calculated by Monte Carlo (noisy lines) assuming the jump probability $\mathcal{P}(z)$ in the form (2.13) with $\gamma = 0.76$ for an infinite crystal and holes generated at $z = 0$. Smooth lines are obtained by numerical evaluation of stable distribution (2.16) of same index.

(b) Similar results for $\gamma = 3.5$. The displayed Monte Carlo results are very close to Gaussian curves $p(z, N)$ of width $\langle z^2 \rangle = N \times (8/15) z_{min}^2$, evaluated according to the central limit theorem.

Comparison of Eqs. (2.13) and (2.19) illustrates a major property of the random walk with a heavy-tailed jump distribution: at large distances, the dominant contribution to $p(z,N)$ results from the single jumps from the starting point – with a pre-factor corresponding to the number of attempts. This result justifies the "on the spot approximation" (OTSA) that was mentioned in Introduction and will be put to a practical use in Sect. 5.

*2d. Stationary hole distribution for constant excitation.*

In the preceding Section 2c we calculated the distribution of holes $p(z,N)$ after a given number of jumps $N$ upon their localized excitation, as if the hole jumps started all at once. We are now concerned with a *continuous* constant excitation and the resultant *stationary* distribution $p_{st}(z)$. To bridge these two problems, we consider the time evolution of the distribution $p(z,t)$ after a short excitation pulse at $t=0$. Given the $p(z,t)$, we can evaluate $p_{st}(z)$ by applying the Duhamel principle [29], viz.

$$p_{st}(z) = \int_0^\infty p(z,t) \exp(-t/\tau) \, dt/\tau, \qquad (2.20)$$

where the exponential factor accounts for the decline of the hole concentration in time, with $\tau \approx \tau_{nr}$ being the average lifetime.[3]

Equation (2.20) suggest a Monte Carlo approach to evaluating $p_{st}(z)$. Firstly, we interpret the simulated distribution $p(z,N)$ as $p(z,t)$ for a fixed "discrete" time $t = N\tau_{rad}$ in units of the radiation emission time. Next, we average the simulated distributions over the durations of random walk $t$ distributed as $\rho(t) = \tau_{nr}^{-1} \exp(-t/\tau_{nr})$. This is equivalent to averaging over the number of jumps (recycling events) $N$ for a given mean value $\overline{N}$, which coincides by definition with the recycling factor $\Phi \equiv \overline{N}$.

$$p_{st}(z)\big|_\Phi = \sum_{N=1}^\infty p(z,N) \exp(-N/\Phi). \qquad (2.21)$$

For $\Phi \gg 1$, the main contribution to the sum for all $z > z_{min}$ comes from terms with $N \gg 1$ and hence both Eqs. (2.20) and (2.21) give very close results.

---

[3] For the ordinary diffusion (Brownian random walk) one has $p(z,t) = (4\pi Dt)^{-1/2} \exp(-z^2/4Dt)$. In this case, Eq. (2.20) yields

$$p_{st}(z) = (2L_D)^{-1} \exp(-|z|/L_D),$$

which describes an exponential decline of the concentration away from the excitation point, with a characteristic length $L_D = \sqrt{D\tau}$, called the diffusion length.

The stationary distribution for a given recycling factor $\Phi$ can also be found in an analytical form by substituting into (2.21) – instead of the Monte Carlo simulated distributions $p(z,N)$ – the quadrature expression (2.16) for the stable distribution. This gives

$$p_{st}(z)\big|_\Phi = \int_0^\infty \frac{\cos(kz)\,dk}{1+\Phi(kz_C)^\gamma}. \qquad (2.22)$$

Equation (2.22) is a new result that generalizes the stationary distribution for a Brownian ($\gamma > 2$) random walk to the case of a Lévy flight ($\gamma < 2$). Analysis of Eq. (2.22) readily shows the existence of two regions in the hole concentration profile, that of asymptotic decay and that of short jumps. In the asymptotic region the hole concentration is formed by repeated one-pass long-distance flights and hence it drops off like the jump probability itself, $p_{st}(z) \propto \mathcal{P}(z) \propto 1/z^{1+\gamma}$. The asymptotic region corresponds to $z \gg \Phi^{1/\gamma} z_C$. The length $z_F \equiv \Phi^{1/\gamma} z_C$ characterizes the front spread distance; in our samples $z_F \gg z_C$, since $\Phi \gg 1$ and $\gamma < 1$. In the short-jumps region, $z \ll z_F$, the concentration of holes drops off with distance at a much slower rate, $p_{st}(z) \propto 1/z^{1-\gamma}$.

The Monte Carlo results for $p_{st}(z)$ are presented in Figs. 2.4 (a, b). We consider a relatively thin InP wafer with optical excitation near the front surface. The simulated distributions drop off with the distance in a non-exponential way, similar to Eq. (2.22) and characteristic of the LF transport. However, details are sensitive to the boundary conditions on the back surface.[4] These effects are illustrated in Fig. 2.4(a), which plots $p_{st}(z)$ for a sample with a particular carrier concentration ($N_D = 6.3 \times 10^{18}$ cm$^{-3}$) but different boundary conditions. A particular strong effect is produced by the *reflecting back* boundary conditions. The corresponding stationary distribution, $p_{rb}(z)$, decays with a much slower exponent compared to that for a semi-infinite medium (non-reflecting back surface). The reflecting boundary conditions are ruled out by our experiments (Sect. 3) although on the first glance they appear plausible, owing to the complete internal reflection of most light rays at the back surface of InP wafer. However, these rays go outside the observation and are effectively extinct.

The simulated distributions $p_{st}(z)$ for a set of differently doped samples and non-reflecting back boundary conditions are presented in Fig. 2.4(b) on a log-log scale. Parameters of the jump probability $\mathcal{P}(z)$ and the recycling factor $\Phi$ are listed in Table I. For comparison, we also show an exponentially decaying distribution corresponding to a Brownian random walk (see footnote[3]) with an exemplary diffusion length $L_D = 45\,\mu\text{m}$.

---

[4] The normalized distributions are found to be practically insensitive to the boundary conditions on the *front* surface. For example, the stationary distributions $p_{rf}(z)$ obtained with the *reflecting front* boundary conditions satisfy for $z > 0$ the expected relation $p_{rf}(z) \approx 2 p_\infty(z)$ with the distribution $p_\infty(z)$ corresponding to fully infinite media ($-\infty < z < \infty$). The factor of 2 corresponds to the contribution of an "image" source of photons provided by the reflecting boundary at $z = 0$.

With the non-reflecting boundary conditions on the back surface, the simulated distributions are accurately described by Eq. (2.22). For moderately doped samples, one can estimate $z_F \approx 300\,\mu m$ and hence the onset of the asymptotic range is at distances larger than the sample thickness $d$. Neither the condition $d \ll z_F$ nor $d \gg z_F$ applies experimentally. In the sample range $0 < z < d$, the stationary distributions have an intermediate asymptotic $p_{st}(z) \propto 1/z^{1+\tilde{\gamma}}$ with $\tilde{\gamma} \approx 0.12 < \gamma$. The concentration extends over a much larger region than could be expected from an exponential distribution and it drops off slower than the single-jump probability $\mathcal{P}(z)$ given by Eqs. (2.8) and (2.13).

**Table I. Parameters of the jump distribution $\mathcal{P}(z)$ and the recycling factor $\Phi$**

| $N_D$ ($10^{18}$ cm$^{-3}$) | 0.2 to 0.6 | 2 | 3.7 | 6.3 | 8 |
|---|---|---|---|---|---|
| $\gamma$ | 0.69 to 0.64 | 0.79 | 0.7 | 0.64 | 0.69 |
| $z_{min}$ (μm) | 0.1 | 0.6 | 0.7 | 1.0 | 1.4 |
| $\Phi$ | 90 | 34 | 19 | 11 | 8 |

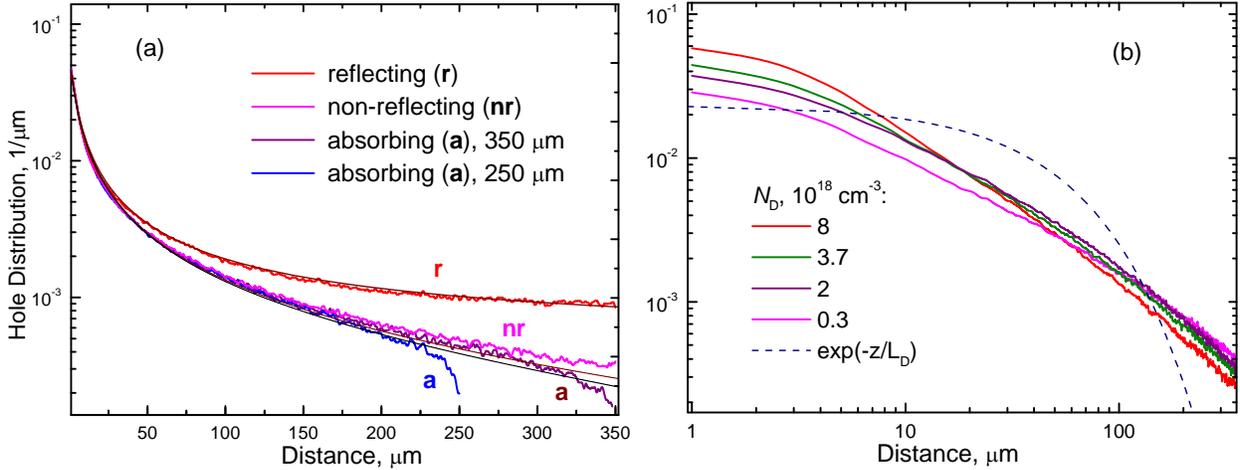

**Fig. 2.4.** Stationary distribution $p_{st}(z)$ of holes generated near the front surface and calculated by Monte Carlo assuming the jump probability $\mathcal{P}(z)$ in the form (2.13). Relevant distribution parameters for differently doped samples are listed in Table I.

(a) InP sample ($N_D = 6.3 \times 10^{18}$ cm$^{-3}$) of finite thickness and different boundary conditions at the back surface, **r**: reflecting, **a**: absorbing, **nr**: non-reflecting (the latter means a semi-infinite sample with no back surface). Dashed lines show the power-law fitting with $p_{st}(z) = c/(z+z_{min})^{1+\tilde{\gamma}}$ where $\tilde{\gamma} = 0.12$.

(b) Distributions $p_{st}(z)$ for differently doped samples with non-reflecting boundary conditions. The dashed line shows the exponential distribution for $L_D = 45\,\mu m$ (see footnote[3]).

Finally, we remark that the one-dimensional Lévy-flight transport is fully described by the integro-differential equation (2.7). This equation, with appropriate boundary conditions, can be solved numerically, using available COMSOL software. We evaluated the hole distributions in this way and found excellent agreement with the Monte Carlo results, except within a region of the order of the hole diffusion length near the sample surface. Far away from that region, the integro-differential equation admits of an analytic solution [5], which is again Eq. (2.22).

## 3. Transmission and reflection luminescence spectra

The basic experiment is illustrated in Fig. 3.1. Thin *n*-type InP wafers (mostly 350 μm, but some further thinned down to 250 μm and 50 μm) were illuminated by short wavelength radiation to ensure short penetration of the incident radiation into the wafer, so that the resulting distribution of holes is dominated by the carrier kinetics. The spectra were taken at room temperature in both reflection and transmission geometries. We shall denote these spectral functions by $I_{\text{refl}}(E)$ and $I_{\text{trans}}(E)$, and the total integrated luminescence intensities by $L_{\text{refl}}$ and $L_{\text{trans}}$, respectively. The setup is schematically shown in Fig. 3.1 (a). It is convenient for the analysis to deal with the ratios of luminescence signals rather than the signals themselves, because the internal reflection factors at the wafer/air interface cancel out, being equal at the front and the back surface.

Figure 3.1 (b) shows the measured intensity ratio of the transmission and reflection luminescence as a function of the majority-carrier doping $N_{\text{D}}$. The ratio $L_{\text{trans}}/L_{\text{refl}}$ increases both at high $N_{\text{D}}$ (due to the Moss-Burstein effect) and at low $N_{\text{D}}$ because of the enhanced photon recycling effect (higher recycling factor $\Phi$).

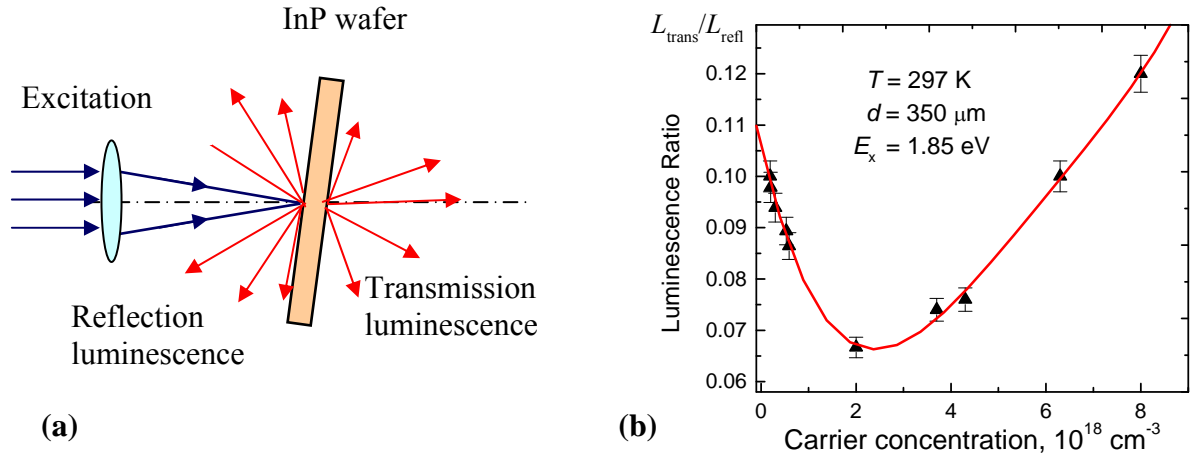

**Fig. 3.1.** Transmission and reflection luminescence spectra.

(a) Schematic experimental setup. The excitation wavelength is chosen short enough to ensure small penetration of the incident radiation.

(b) Intensity ratio of the transmission and reflection luminescence as function of the doping $N_{\text{D}}$.

The measured spectra $I_{refl}(E)$ and $I_{trans}(E)$ are distorted compared to the intrinsic emission spectrum $\mathcal{N}(E)$ in a different way, because of the different reabsorption-filtering geometry,

$$I_{trans}(E) = F_{trans}(E)\mathcal{N}(E), \tag{3.1a}$$

$$I_{refl}(E) = F_{refl}(E)\mathcal{N}(E). \tag{3.1b}$$

The filtering factors $F_{refl}(E)$ and $F_{trans}(E)$, are expressed through one-pass filtering functions,

$$F_1(E) = \int_0^d p(z)\exp[-\alpha(E)z]\,dz, \tag{3.2a}$$

$$F_2(E) = \int_0^d p(z)\exp[\alpha(E)(d-z)]\,dz, \tag{3.2b}$$

where $p(z)$ is the non-equilibrium stationary hole concentration that results from the steady-state excitation near the front surface $(z=0)$ of the InP wafer of thickness $d$. Taking into account multiple reflections at surfaces of the wafer, this expression is of the form [14],

$$F_{trans}(E) = (1-R)\frac{F_2 + RF_1\exp(-\alpha d)}{1-R^2\exp(-2\alpha d)}, \tag{3.3a}$$

$$F_{refl}(E) = (1-R)\frac{F_1 + RF_2\exp(-\alpha d)}{1-R^2\exp(-2\alpha d)}, \tag{3.3b}$$

where $R \approx 0.3$ is the InP reflection coefficient.[5]

It is possible *in principle* to experimentally determine both factors $F_{refl}(E)$ and $F_{trans}(E)$ from precisely measured spectra $I_{refl}(E)$, $I_{trans}(E)$ and $\alpha(E)$ [the latter determines $\mathcal{N}(E)$ by the VRS relation (2.10)]. Next, one could invert Eqs. (3.3) to determine $F_1(E)$ and $F_2(E)$, and use Eqs. (3.2) as integral equations for $p(z)$. In fact, viewed as functions of the argument $\alpha(E)$ the functions $F_1(\alpha)$ and $F_2(\alpha)$ represent Laplace transformations of a modified function $p(z)$, naturally extended to the infinite domain. Again *in principle*, one can obtain $p(z)$ by inverting these Laplace transformations numerically. However, numerical inversion of the Laplace transform is a classical ill-conditioned mathematical problem and the accuracy of our spectral measurements is not sufficient for finding a meaningful solution. Instead, our approach [14] was to calculate the reabsorption-filtering functions from the model distribution of holes in the layer, and compare the results with the functions found from the experiment.

---

[5] In numerical calculations [14], we took into account the measured dependence $R(E)$ and its variation with $N_D$.

The key experimental function for our analysis is the *ratio* of the transmission and reflection spectra, which in light of Eqs. (3.1) and (3.30 is given by

$$\mathcal{R}(E) \equiv \frac{I_{\text{trans}}(E)}{I_{\text{refl}}(E)} = \frac{F_2 + RF_1 \exp(-\alpha d)}{F_1 + RF_2 \exp(-\alpha d)} \ . \tag{3.4}$$

This ratio has important advantages for the analysis of the spatial hole distribution $p(z)$, firstly because it does not depend on details of the intrinsic emission spectrum. Furthermore, it is not sensitive to multiple reflections, since the denominators of Eqs. (3.3) cancel out. At the same time, it is quite sensitive to $p(z)$ through $F_1(E)$ and $F_2(E)$. Therefore, the ratio $\mathcal{R}(E)$ is well suited to quantify the spatial hole distribution. Figure 3.2 shows the ratio for several samples, lightly doped (a) and heavily doped (b).

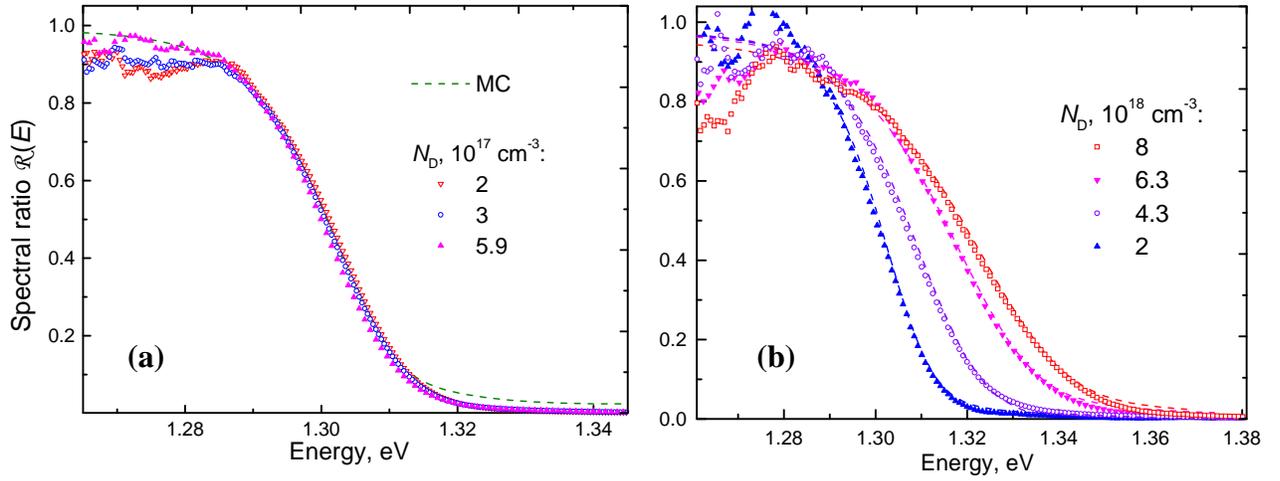

**Fig. 3.2.** Ratio $\mathcal{R}(E)$ of transmission and reflection luminescence spectra measured in 350 μm thick InP wafers of various doping levels $N_D$. For several select $N_D$ we also show theoretical curves, calculated with Eqs. (3.4) and (3.2) with Monte Carlo modeled model distributions $p(z)$ as in Fig. (2.4b).

(a) Lightly doped samples; (b) heavily doped samples.

The experimental ratio curves show nearly perfect fit to the theoretical curves calculated from Eq. (3.4) with $F_1(E)$ and $F_2(E)$ given by Eqs. (3.2), where we take for $p(z)$ the model stationary distributions $p_{\text{st}}(z)$ evaluated by Monte Carlo or by Eq. (2.22). The only adjustable parameter is the recycling factor, which was previously estimated independently from time-resolved luminescence kinetics [25]. The agreement is excellent.

Nevertheless, this fit, however perfect, does not provide an unambiguous evidence of Lévy flight transport. The "problem" is that we can obtain a reasonable fit also by assuming an exponential distribution (see footnote[3]) with $L_D$ as an adjustable parameter [14]. For example, the $N_D = 2 \times 10^{18}$ cm$^{-3}$ data for $\mathcal{R}(E)$ are well fit with $L_D = 45\,\mu m$, cf. the exponential curve in Fig. 2.4 (b), corresponding to the Brownian random walk.

An unambiguous experimental demonstration of LF transport is presented in Fig. 3.3, where we plot the ratio $\mathcal{R}(E)$ for samples of same doping but different thickness $d$. In this case, the theoretical fit based on distributions $p_{st}(z)$ evaluated for the Lévy flight remains nearly perfect, whereas the exponential distribution fails miserably. The exponential approximation fails to describe thin and thick samples simultaneously. To fit the data for 50 μm sample, one would have to assume an $L_D$ substantially shorter than the sample thickness. This, however, would be in contradiction with the fairly high intensity of transmitted radiation in the 350 μm sample of the same doping. The power-law decay of the hole concentration is steep enough at short distances (steeper than an exponent) to fit the data for the thin sample, and at the same time slow enough at large distances (again, compared to an exponent) to account for thick samples.

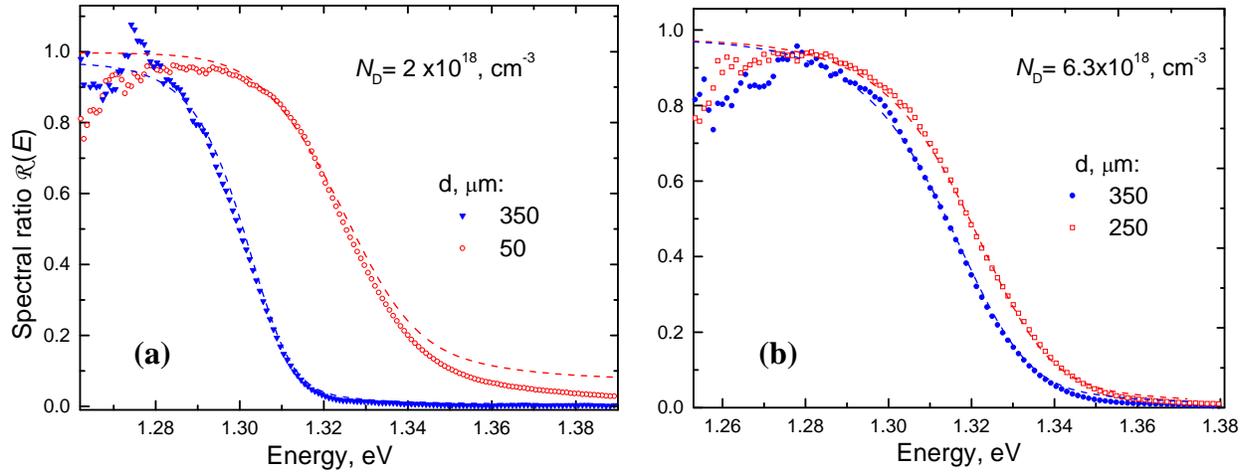

**Fig. 3.3.** Ratio $\mathcal{R}(E)$ of transmission and reflection luminescence spectra measured in a pair of thick and thin InP wafers of same doping concentration $N_D$. Dashed lines show theoretical curves, calculated with Eqs. (3.4) and (3.2) with Monte Carlo modeled model distributions $p(z) = p_{st}(z)$ as in Fig. (2.4b).

(a) Moderately doped samples, 350 μm and 50 μm thick;
(b) Heavily doped samples, 350 μm and 250 μm thick.

## 4. Luminescence filtering and Urbach tails

Luminescence studies described in Sect. 3 shed light on the anomalous transport properties over distances limited by the sample thickness, in our case < 350 μm. To circumvent this limitation, photoluminescence experiments were performed [15], where the luminescence spectra were excited by a red laser in a narrow spot on the broadside surface of an InP wafer but registered from the edge of the wafer, a distance $d$ away from the excitation spot.

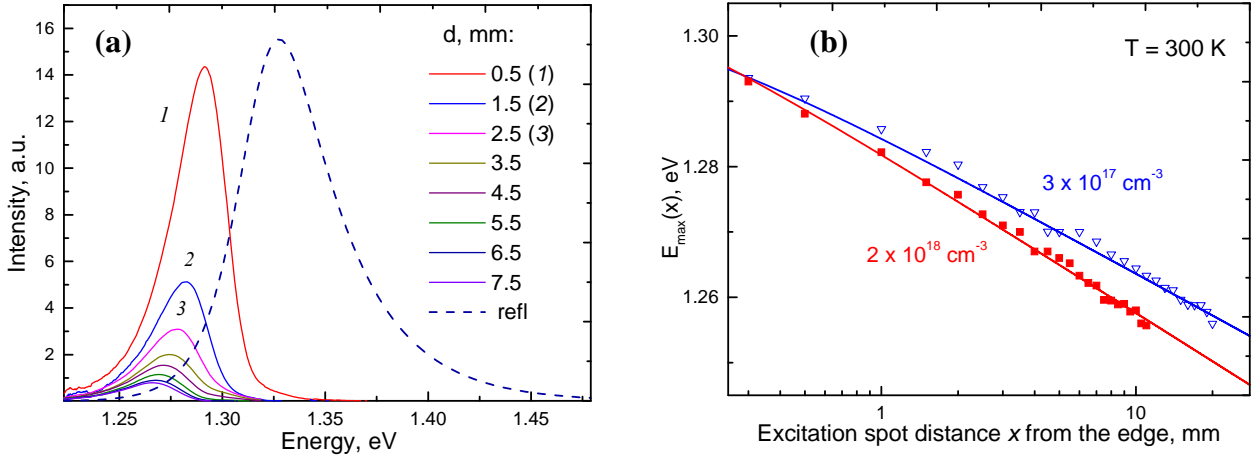

**Fig. 4.1.** Edge luminescence experiment [15].

(a) Room-temperature luminescence spectra for $n$-InP sample, $(3\times10^{17}\text{ cm}^{-3})$ observed at increasing distances $d$ between the excitation spot and the wafer edge; the dashed line shows the emission spectrum;

(b) Shift of the luminescence peak $E_{\max}(d)$ for two samples of different doping.

The observed spectra, Fig. 4.1, show two interesting features. Firstly, we see a power-law decrease of the integrated intensity and, secondly, a noticeable red shift of the spectral maximum. The power law is indicative of the anomalous transport but the exact Lévy-flight exponent is hard to extract from this experiment because of the irregular geometry producing specks of reflection. The red shift, on the other hand, can be analyzed very accurately and relate to the important spectral parameters, viz. the Urbach tails $\Delta$ and $\Delta'$ corresponding to absorption and emission spectra, respectively, and defined by their behavior deep in the red wing,

$$\alpha(E) = \alpha_0 \exp\left(\frac{E-E_{\mathrm{G}}}{\Delta}\right), \tag{4.1a}$$

$$S(E) = S_0 \exp\left(\frac{E-E_{\mathrm{G}}}{\Delta'}\right). \tag{4.1b}$$

Both tailing parameters depend on the doping concentration and the temperature, $\Delta = \Delta(T, N_D)$ and $\Delta' = \Delta'(T, N_D)$. In our $N_D = 3 \times 10^{17}$ cm$^{-3}$ InP samples at room temperature, $\Delta = 9.4$ meV and $\Delta' = 15$ meV. For $N_D = 2 \times 10^{18}$ cm$^{-3}$, we have $\Delta = 10.6$ meV and $\Delta' = 16$ meV. The pre-exponential absorption factor $\alpha_0$ cannot be measured independently of the small uncertainty in the bandgap $E_G$, but it should not depend on the doping and the value $\alpha_0 = 1.1 \times 10^4$ cm$^{-1}$ fits very well in a wide temperature range for undoped samples, where $E_G$ is known accurately.

The observed red shift, for several temperatures, fits very accurately to the expression,

$$E_{max}(d) = E_G - \Delta \ln\left(\frac{\alpha_0 (d + d_{min})}{a}\right). \tag{4.2}$$

This expression comprises two empirical parameters, $d_{min}$ and $a$. The former of these, reflects details of the experimental geometry (finite width and depth of the excitation spot) and, since $d_{min} < 200$ μm for all samples, it is of no importance when distances $d$ are in the range of 1 to 20 mm, i.e., for $d \gg d_{min}$. The main empirical content of the $E_{max}(d)$ dependence resides in the parameter $a = a(T, N_D)$. For $N_D = 3 \times 10^{17}$ cm$^{-3}$ and $N_D = 2 \times 10^{18}$ cm$^{-3}$ at $T = 300$ K, we have respectively $a = 0.63$ and $a = 0.68$.

The observed $E_{max}(d)$, including the values of $a$, can be reproduced in a simple model that attributes the luminescent peak shift to wavelength-dependent filtering of outgoing radiation by the sample absorption. In the spirit of the on-the-spot approximation, the observed radiation at the edge arises from repeated emissions at the same spot where the initial hole was generated. Therefore, we assume that the position of the peak observed at distance $d$ from the excitation spot is determined by the transparency of InP wafer to the emission spectrum $S(E)$. In other words, the observed edge spectrum near its maximum is described by

$$S_{obs}(E, d) = S(E) \times \exp[-\alpha(E)d]. \tag{4.3}$$

The strong refraction of outgoing radiation and a relatively small observation angle ensure a small and constant range of the angles of incidence. Therefore, the $d$ dependence corresponds to one-dimensional attenuation of light. The maximum of the observed spectrum can be found from the expression $dS_{obs}(E,d)/dE = 0$, which, in light of (4.3), takes the form

$$\left.\frac{d \ln[S(E)]}{dE}\right|_{max} = d \times \left.\frac{d\alpha(E)}{dE}\right|_{max}, \tag{4.4}$$

Substituting Eqs. (4.1) into (4.4), we find an expression of the form (4.2) with the parameter $a$ given by $a = \Delta/\Delta'$. For $N_D = 3 \times 10^{17}$ cm$^{-3}$ and $N_D = 2 \times 10^{18}$ cm$^{-3}$, respectively, the Urbach tail ratio gives $a = 0.63$ and $a = 0.67$ in a remarkable agreement with the above empirical values. Similar agreement is obtained for other samples at all temperatures [15].

For all studied cases, the values of $\Delta$ obtained from the slope of $\ln\alpha(E)$ and the slope of the dependence of $E_{\max}(\ln d)$ are very close, the difference never exceeding 0.2 meV. Thus, the described edge luminescence method provides an independent way of measuring the tailing parameters. This method can be indispensable (in fact, the only available) in the case when the residual absorption is strong.

Edge luminescence studies lend further support to the on-the-spot approximation (OTSA) that was justified theoretically in Sect. 2c: at large distances, the dominant contribution to the observed spectra results from repeated jumps from the starting point. This principle will be used in the evaluation of photon collection efficiency in semiconductor scintillators, Sects. 4 and 5.

## 5. Photon collection efficiency in InP scintillator

The key issue in implementing a semiconductor scintillator is to make sure that photons generated deep inside the semiconductor slab could reach its surface without tangible attenuation. However, semiconductors are usually opaque at wavelengths corresponding to their radiative emission spectrum. Our group has been working on the implementation of scintillators based on direct-gap semiconductors. For the exemplary case of InP, the luminescence spectrum is a band of wavelengths near 920 nm. The original idea [30] was to make InP relatively transparent to this radiation by doping it heavily with donor impurities, so as to introduce the Burstein shift between the emission and the absorption spectra.

Here we shall describe another approach [16], based on the photon recycling effect owing to the high radiative efficiency of best direct-gap semiconductors, such as InP. In these materials, an act of interband absorption does not finish off the luminescent photon; it merely creates a new minority carrier and then a new photon in a random direction. The resultant random walk has been described (Sect. 2) as the Lévy flight of holes (or photons).

Consider an InP scintillator slab with two photoreceiver systems integrated on the opposite sides of the slab [31]. Exemplarily, these are epitaxially grown InGaAsP photodiodes [32]. Let the interaction occur a distance $z$ from the detector top surface, as indicated in Fig. 5.1, producing minority carriers (holes). A hole has the probability $\eta$ (the radiative efficiency, Eq. 2.1) to generate a photon (distributed in energy according to the emission spectrum, Eq. 2.10). The generated photon can either reach the detectors (probabilities $\pi_1$ and $\pi_2$, respectively) or disappear through free-carrier absorption (single-pass probability $\pi_{\text{FCA}}$). All these probabilities depend on $z$. The combined probability $\Pi(z) = \pi_1 + \pi_2 + \pi_{\text{FCA}}$ describes the likelihood of the photon loss at this stage and the alternative, $1-\Pi(z)$, is the probability that a new hole is created. The cycle of hole-photon-hole transformation repeats *ad infinitum*. Most of the scintillation reaching the detectors' surface are not photons directly generated at the site of the gamma particle interaction, but photons that have been re-absorbed and re-emitted a multiple number of times.

The detector signals $D_1$ and $D_2$ add single-pass contributions from different cycles. The sum can be found as geometric progression, giving (per unit strength of initial excitation)

$$D_i(z) = \eta\,\pi_i(z) \times \sum_{n=0} \left[\eta(1-\Pi)\right]^n = \frac{\eta\,\pi_i(z)}{(1-\eta)+\eta\,\Pi(z)}, \tag{5.1}$$

where $i = 1, 2$. Taking into account Eq. (2.1), the photon collection efficiency, $\text{PCE} \equiv D_1 + D_2$ is given by

$$\text{PCE} = \frac{\pi_1(z)+\pi_2(z)}{[(\tau_{\text{rad}}/\tau_{\text{nr}})+\pi_{\text{FCA}}(z)]+[\pi_1(z)+\pi_2(z)]}. \tag{5.2}$$

We note that for high photon recycling ($\eta \to 1$ and $\pi_{\text{FCA}} \to 0$), the entire luminescence is collected – even though the single-pass probabilities $\pi_1$ and $\pi_2$ may not be high due to interband absorption. The efficiency of photon collection is thus limited by parasitic processes, such as FCA and nonradiative recombination of holes. If these are minimized, one can have an "opaque" but ideal semiconductor scintillator.

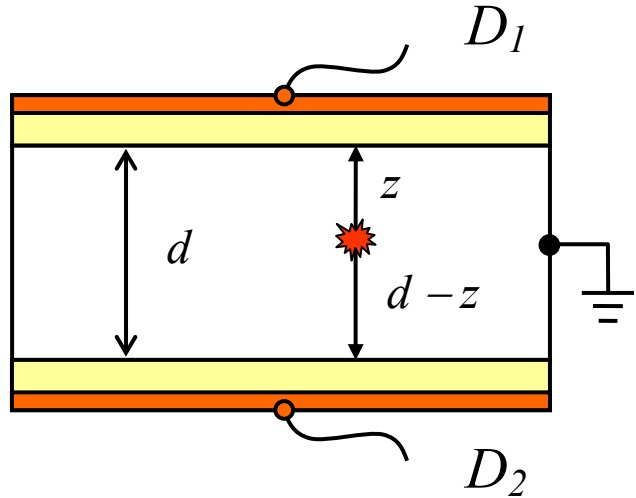

**Fig. 5.1.** Schematic cross-section of InP scintillator with two epitaxial photodiodes grown on both sides. Interaction with incident gamma photon (shown by the explosion symbol) occurs a distance $z$ from the top surface and both photodiode signals depend on this distance.

The only approximation involved in Eq. (5.1) is the assumption that every act of recycling occurs at the same place $z$ where the initial interaction occurred, and therefore the same probabilities $\pi_1(z)$ and $\pi_2(z)$ appear at all stages of the recycling. This "on-the-spot" approximation (OTSA) has reduced the summation of an infinite series to a geometric progression and allowed us to obtain the result in a closed form. The physical motivation for OTSA (the Lévy flight nature of random walk involved in the recycling) was presented in Sects. 2c and 4. Here we remark that going beyond OTSA (by direct Monte Carlo evaluation of PCE) does not change the results qualitatively, although it slightly enhances our estimate of PCE.

The single-pass probabilities $\pi_1(z)$ and $\pi_2(z) = \pi_1(d-z)$ can be evaluated by integration over the isotropic distribution of photon directions and the random energies in the emission spectrum,

$$\pi_1(z) = \int \mathcal{N}(E)\, \pi(z,E)\, dE,$$
$$\pi(z,E) = \int_0^\infty \exp[\alpha_i(E) r]\, \frac{\cos\theta}{2r^2}\, \rho\, d\rho, \tag{5.3}$$

where $\rho = z\tan\theta$ and $r = z\sec\theta$. Similarly, we can evaluate the single-pass probability of free-carrier absorption in terms of the FCA coefficient $\alpha_{fc}$. If we neglect in the exact expression [16] for $\pi_{FCA}(z)$ the corrections due to $\pi_1(z)$ and $\pi_2(z)$, but retain the dominant process of interband absorption $\alpha_i$, then $\pi_{FCA}$ no longer depends on $z$, viz.

$$\pi_{FCA} = \int \mathcal{N}(E)\, \frac{\alpha_{fc}(E)}{\alpha_{fc}(E) + \alpha_i(E)}\, dE. \tag{5.4}$$

Practically, for moderately doped samples, one has $\pi_{FCA} \ll \tau_{rad}/\tau_{nr}$ and FCA can be neglected. The calculated PCE is shown in Fig. 5.2 (a) for three moderate doping concentrations. We see that photon recycling delivers a reasonable fraction of the scintillation to the wafer surface, this fraction being higher for samples with higher radiative efficiency. We note, however, that the photon yield depends on the exact position of the interaction relative to the wafer surfaces. This spells trouble for the needed precise quantification of the energy deposited by a gamma quantum. The problem is how to distinguish the signal arising from a large energy deposited far from the photoreceiver surface from that arising from smaller energy deposited nearby. The problem arises from the attenuation of the optical signal.

However, if we knew the position $z$ of the gamma interaction event, we could correct for the attenuation. The solution [31] is based on tallying the signals $D_1$ and $D_2$ individually. The relative strength of the two signals provides a good measure of event position. A convenient characteristic is the position determining ratio, defined by $\text{PDR} \equiv (D_1 - D_2)/(D_1 + D_2)$ and plotted in Fig. 5.2 (b) as a function of position, $\text{PDR}(z)$. From Eq. (5.1) we find

$$\text{PDR} = \frac{\pi_1(z) - \pi_2(z)}{\pi_1(z) + \pi_2(z)}. \tag{5.5}$$

We see that the PDR is an excellent measure of $z$. The simultaneous detection by *both* detectors of the scintillation arising from the same interaction event, allows us to determine the position of the event and correct for attenuation.

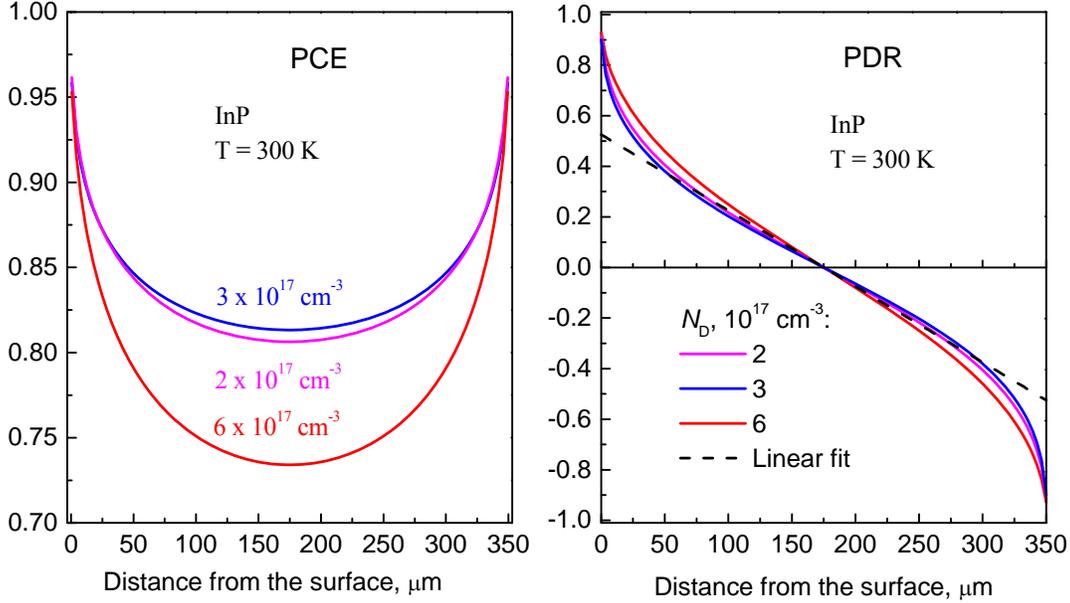

**Fig. 5.2.** Model calculations in the "on-the-spot" approximation for 350 µm thick InP wafers of different doping concentration.

(a) Photon collection efficiency $\mathrm{PCE}\,(z)$ calculated according to Eq. (5.2) as a function of the position $z$ of the interaction

(b) Position determining ratio $\mathrm{PDR}\,(z)$ calculated according to Eq. (5.5) as a function of $z$.

As seen from Eq. (5.2), the efficiency of photon collection in a scintillator based on photon recycling depends strongly on the radiative efficiency of the material and is maximized when $\eta \to 1$, or, equivalently, when $\tau_{\mathrm{rad}}/\tau_{\mathrm{nr}} \to 0$. Since $1/\tau_{\mathrm{rad}} = BN_{\mathrm{D}}$ and $1/\tau_{\mathrm{nr}} = A + CN_{\mathrm{D}}^2$, this ratio is non-monotonic in concentration [17] and has a minimum when $N_{\mathrm{D}}^2 = A/C$. In our series of InP samples, the optimum doping is $N_{\mathrm{D}} = 3 \times 10^{17}\ \mathrm{cm}^{-3}$, where $\eta \approx 98\%$ or even higher [14, 17]. Unfortunately, the high radiative efficiency of the low-doped InP scintillator material does not survive the high-temperature treatment involved in the epitaxial growth and processing of the quaternary InGaAsP *pin* diodes [32]. The bright luminescence of the virgin wafers degrades by nearly two orders of the magnitude upon the heat treatment. The degradation is thermally activated and is apparently related to defects inherent in a bulk Czochralski-grown wafer.

We have been therefore led to explore the possibility of all-epitaxial scintillator. It is known that the luminescent properties of low-doped epitaxial layers, as opposed to those of a bulk wafer, *do not* degrade under high-temperature treatment. Thick (e.g., millimeter-thick) free-standing layers can be grown by such epitaxial techniques as HVPE ("Hydride Vapor Phase Epitaxy," a growth technique with rates exceeding 100 µm an hour [33]) and can be expected to have superior non-degrading luminescence properties. While we began to pursue such an approach experimentally, we explored theoretically the possibilities it offers. The most recent advance in this regard was our invention [18] of an artificially layered scintillator material that comprises alternating thick wide-gap "barrier" and thin narrow-gap "well" layers. The wells constitute the radiation sites that are *pumped by light* generated in the barriers. The idea is discussed in the next Section.

## 6. Layered scintillator based on photon-assisted transport of holes to radiation sites

It would be attractive to implement a structure that could be scaled to thicknesses $d \approx 1$ mm and higher. We decided to explore the possibilities that would arise with an all-epitaxial fast-growth technique. This has led to our invention [18] of a new scintillator principle, described below.

The inventive scintillator material illustrated in Fig. 6.1 is a direct bandgap semiconductor heterostructure that comprises alternating thick wide-gap "barrier" layers $B_j$ and thin narrow-gap "well" layers $W_j$. Exemplarily, we take the wide-gap layers as low-doped InP and the well layers as lattice-matched InGaAsP alloy of 100 meV narrower bandgap. The assumed absorption and emission spectra shown in Fig. 6.2 are based on the experimental spectra of Fig. 2.1.

Minority carriers generated in the wide-gap material by an incident high-energy particle, recombine there radiatively, producing *primary* scintillation light that is captured by the narrow-gap wells generating new minority carriers therein. Recombination of these new minority carriers in the narrow-gap wells generates *secondary* longer wavelength scintillation – to which the entire layered structure is largely transparent. It is important that the separation between narrow-gap wells *is not limited by the minority-carrier diffusion length* and can be as large as hundreds of microns.

In the conventional scintillator language, the narrow-gap wells can be viewed as radiation sites that emit light at subband wavelengths [30, 34]. However, no travel of carriers to these radiation sites is now contemplated. This is a radical departure from all prior-art activated scintillators, where charge carriers were supposed to travel to the radiation sites and the distance to travel had to be minimized by increasing their concentration. The finite travel distance, even when minimized, leads to an unwelcome "non-proportionality" [35] of activated dielectric scintillators, which impedes their energy resolution. The present invention circumvents this requirement.

The main advantage of the inventive scintillator structure is the possibility to enhance the overall thickness of the semiconductor body beyond 1 mm.

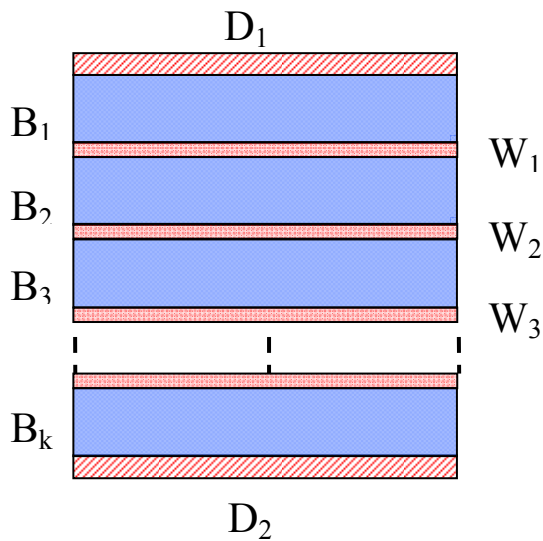

**Fig. 6.1**. All-epitaxial multilayered scintillator [18]. The scintillator body comprises a sequence of $k$ barrier (InP) layers $B_i$ of thickness $b$, alternating with $k-1$ well layers $W_i$ of thickness $w$ made of quaternary InGaAsP lattice-matched to InP.

Exemplarily, $k = 10$, $b = 100$ μm and $w = 1$ μm, so there are altogether 10 barriers and 9 wells of total thickness about 1 mm.

The wells collect the primary photons generated in the barriers and in turn generate secondary photons – the scintillation output.

The structure includes photoreceivers $D_1$ (top) and $D_2$ (bottom), sensitive to the scintillation.

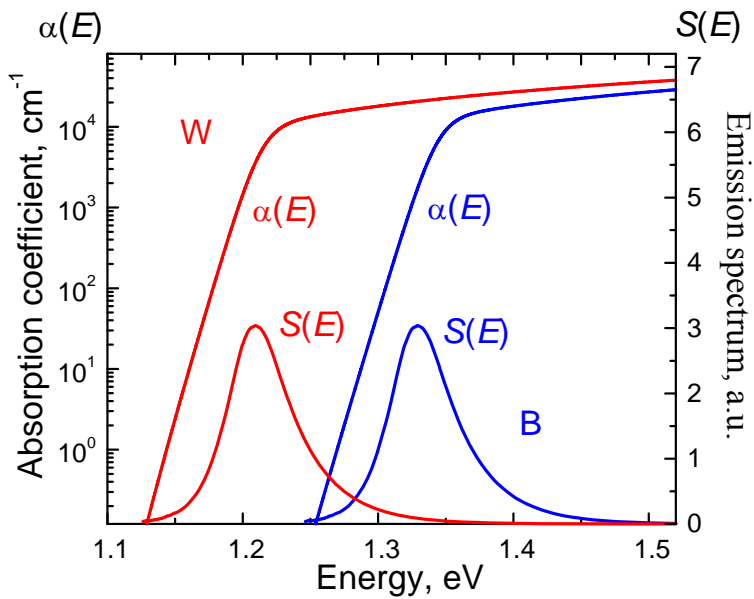

**Fig. 6.2.** Absorption and emission spectra of the multilayer structure, assumed in our calculations. For barrier layers B, the assumed spectra coincide with experimental curves (Fig. 2.1) for moderately doped InP. For the well layers W, the assumed curves describe similarly-doped lattice-matched InGaAsP alloy whose bandgap is 100 meV narrower than that of InP.

Figure 6.3 shows the results of calculating the response of the 1 mm thick multilayered structure described by Figs. 6.1 and 6.2. Calculations presented are based on the absorption and emission spectra of InP and take full account of the anomalous transport properties of photons in the high-radiative efficiency InP material.

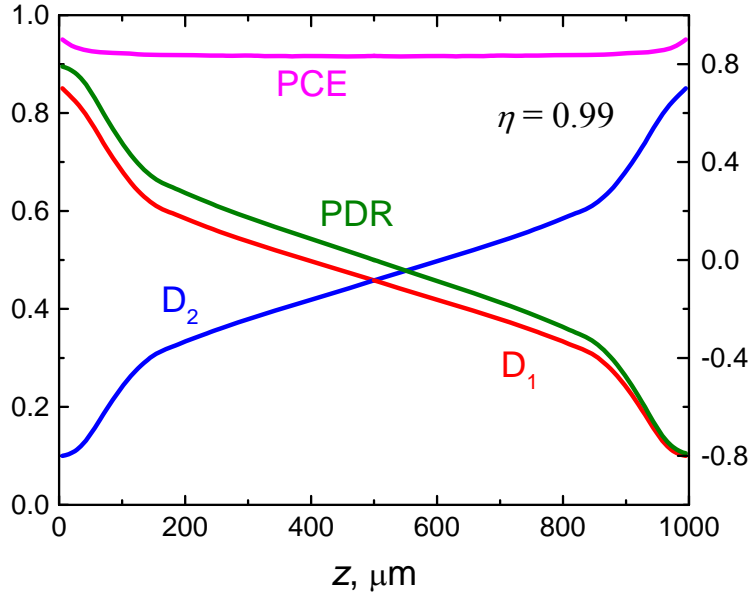

**Fig. 6.3.** Normalized photodiode signals $D_1$ and $D_2$, the photon collection efficiency (PCE = $D_1 + D_2$), and the position-determining ratio PDR = $(D_1 - D_2)/(D_1 + D_2)$ calculated for the multilayered structure described by Figs. 6.1 and 6.2 as a function of the position $z$ of the interaction, counted from the bottom photoreceiver plane. The assumed quantum radiative efficiency is $\eta = 99\ \%$ throughout the structure.

Photon-assisted transport enables us to space the radiation sites by a large distance, much larger than the diffusion length of carriers. Ultimately, this may lead to the implementation of centimeter-thick semiconductor scintillators. The capability for vertical position determination by the PDR function makes thick scintillators very attractive for their desired application [17] as a voxel (3D pixel) in a three-dimensional array of radiation detectors. The PDR resolves the $z$ coordinate to a much finer degree than the linear dimension of the voxel in the $z$ direction and enables one to replace a 3D stack of 2D arrays by a single 2D array of thick layered scintillators.

## 7. Conclusion

We have discussed a remarkable laboratory system for studying Lévy flights, namely high-radiative-efficiency semiconductors. The Lévy-flight model provides an adequate description for the photon-assisted transport of minority carriers, where photons are repeatedly recycled by the interband absorption/re-emission processes. Based on photoluminescence experiments in high-efficiency bulk InP, we have demonstrated an unambiguous evidence for the Lévy-flight nature of this anomalous non-diffusive transport.

We have developed a quantitative theoretical description of the steady-state distributions of minority carriers that emerge in the anomalous transport. The description, based on the mathematics of the *stable* Lévy-flight distributions, has been checked against both the experiments and Monte Carlo simulations. Our key experiments involve spectral *ratios* of the luminescence observed in the transmission and reflection geometries. Particularly revealing are ratios obtained with the variable wafer thickness.

Supported by the Lévy-flight model, we have formulated the so-called *on-the-spot approximation* (OTSA) that proves to be very useful in describing photon collection processes in semiconductor scintillators based on photon recycling. Equations of the OTSA demonstrate the tantalizing possibility of implementing a direct-gap semiconductor scintillator that is *opaque* in the usual sense at the wavelength of its own scintillation. Nevertheless, this scintillator will have nearly ideal photon collection efficiency.

Finally, we discussed our recent invention, not yet reduced to practice, of a *layered* semiconductor scintillator, in which the Lévy-flight photon-assisted transport is used to deliver the generated primary minority carriers to the "radiation sites" implemented as semiconductor wells of narrower bandgap, spaced apart by distances much larger than the diffusion length of the minority carriers. The nearly ideal characteristics of the layered millimeter-thick scintillator can be scaled up to the dimensions of several centimeters.

**Acknowledgements**

This work was supported by the Domestic Nuclear Detection Office (DNDO) of the Department of Homeland Security, by the Defense Threat Reduction Agency (DTRA) through its basic research program, and by the New York State Office of Science, Technology and Academic Research (NYSTAR) through the Center for Advanced Sensor Technology (Sensor CAT) at Stony Brook.